\documentclass[prd,twocolumn,amsmath,amssymb,nofootinbib,floatfix,superscriptaddress]{revtex4}

\usepackage{graphicx,bm,float}

\makeatletter
\def\graphicscale{\twocolumn@sw{0.3}{0.4}}
\def\graphicthreescale{\twocolumn@sw{0.3}{0.4}}

\begin{document}

\title{Asymptotic low-temperature critical behavior of \\
two-dimensional multiflavor lattice SO($N_c$) gauge theories}

\author{Claudio Bonati} 
\affiliation{Dipartimento di Fisica dell'Universit\`a di Pisa 
       and INFN, Pisa, Italy}

\author{Alessio Franchi} 
\affiliation{Dipartimento di Fisica dell'Universit\`a di Pisa,
       Pisa, Italy}

\author{Andrea Pelissetto}
\affiliation{Dipartimento di Fisica dell'Universit\`a di Roma Sapienza
        and INFN, Roma, Italy}

\author{Ettore Vicari} 
\affiliation{Dipartimento di Fisica dell'Universit\`a di Pisa
       and INFN, Pisa, Italy}

\date{\today}

\begin{abstract}
We address the interplay between global and local gauge nonabelian
symmetries in lattice gauge theories with multicomponent scalar
fields.  We consider two-dimensional lattice scalar nonabelian gauge
theories with a local SO($N_c$) ($N_c\ge 3$) and a global O($N_f$)
invariance, obtained by partially gauging a maximally O($N_f
N_c$)-symmetric multicomponent scalar model. Correspondingly, the
scalar fields belong to the coset $S^{N_f N_c-1}$/SO($N_c$), where
$S^N$ is the $N$-dimensional sphere.  In agreement with the
Mermin-Wagner theorem, these lattice SO($N_c$) gauge models with
$N_f\ge 3$ do not have finite-temperature transitions related to the
breaking of the global nonabelian O($N_f$) symmetry. However, in the
zero-temperature limit they show a critical behavior characterized by
a correlation length that increases exponentially with the inverse
temperature, similarly to nonlinear O($N$) $\sigma$ models.  Their
universal features are investigated by numerical finite-size scaling
methods.  The results show that the asymptotic low-temperature
behavior belongs to the universality class of the two-dimensional
RP$^{N_f-1}$ model.
\end{abstract}

\maketitle


\section{Introduction}
\label{intro}

Lattice gauge models provide effective theories in various physical
contexts, ranging from fundamental
interactions~\cite{Wilson-74,Weinberg-book} to emerging phenomena in
condensed matter physics~\cite{Sachdev-19,Anderson-book}. They provide
mechanisms for fundamental phenomena, such as confinement and the
Higgs mechanism, which explain the spectrum of subnuclear systems
interacting via strong and electroweak forces, superconductivity,
etc...  The interplay between global and local gauge symmetries is
crucial to determine the main features of the theory, such as the
nature of the spectrum, the degeneracy of the energy levels, the phase
diagram, the nature and universality classes of their thermal and
quantum transitions.

In the case of two-dimensional (2D) lattice gauge models, the
interplay of non-Abelian global symmetries and local gauge symmetries
determines the large-scale properties of the system in the
zero-temperature limit, and therefore, the statistical field theory
realized in the corresponding continuum limit~\cite{ZJ-book}.  These
issues have been addressed in the multicomponent Abelian-Higgs
model~\cite{BPV-19-ah2}, characterized by a global U($N_f$) symmetry
($N_f\ge 2$) and a local U(1) gauge symmetry, and in the multiflavor
scalar quantum chromodynamics~\cite{BPV-20-qcd2}, characterized by a
global U($N_f$) symmetry and a local SU($N_c$) gauge symmetry. The
results of Refs.~\cite{BPV-19-ah2,BPV-20-qcd2} provide numerical
evidence that the asymptotic low-temperature behavior of these 2D
lattice gauge models always belongs to the universality class of the
2D CP$^{N_f-1}$ field theory~\cite{ZJ-book}. The universality class is
only determined by the global U($N_f$) symmetry of the model. The
local gauge symmetry apparently does not play any role: models with
different gauge symmetry but with the same global invariance have the
same large-scale low-temperature behavior.  These results may be
interpreted as a numerical evidence of a more general
conjecture~\cite{BPV-20-qcd2}: the renormalization-group flow
determining the asymptotic low-temperature behavior is generally
controlled by the 2D statistical field theories associated with the
symmetric spaces~\cite{BHZ-80,ZJ-book} that have the same global
symmetry. This is indeed the case of the Abelian-Higgs model and of
scalar chromodynamics, whose low-temperature behavior is always
controlled by the 2D CP$^{N_f-1}$ field theory.

To achieve additional evidence of the above conjecture, we extend the
analysis to other 2D lattice models, characterized by different global
and local gauge symmetries.  For this purpose, we consider 2D lattice
models with real scalar fields, which are invariant under global and
local gauge transformations that belong to orthogonal groups.  In
particular, we consider lattice gauge models that are invariant under
SO($N_c$) local transformations and under O($N_f$) global
transformations ($N_c$ will be referred to as the number of colors and
$N_f$ as the number of flavors), focusing on the case $N_f\ge 3$, so
that the symmetry group is nonabelian.  According to the Mermin-Wagner
theorem~\cite{MW-66}, these lattice gauge models do not present a
finite-temperature transition associated with the breaking of the
global O($N_f$) symmetry. However, they are expected to develop a
critical behavior in the zero-temperature limit (for $N_f=2$ the
global symmetry group is the abelian group O(2), so that a
finite-temperature Berezinskii-Kosterlitz-Thouless transition is
possible).  We study the universal features of the asymptotic
zero-temperature behavior for $N_f=3,4$ and $N_c=3,4$ by means of
finite-size scaling (FSS) analyses of Monte Carlo (MC) results.
According to the above-mentioned conjecture, the asymptotic behavior
should be that of a statistical theory defined on a symmetric space
with the same global symmetry. We provide a theoretical argument that
shows that the appropriate model is the 2D RP$^{N_f-1}$ model, in
which the fields effectively belong to the real projective space in
$N_f$ dimension, a symmetric space which is invariant under global
O($N_f$) transformations. Note that the associated symmetric space is
not the $N_f$-dimensional sphere, that has the same global symmetry
group.  This is due to the fact that the low-energy behavior is
essentially characterized by a bilinear operator (a projector) that is
invariant under local ${\mathbb Z}_2$ transformations.  We anticipate
that the numerical results confirm the conjecture.

The paper is organized as follows. In Sec.~\ref{model} we introduce
the lattice nonabelian gauge models that we consider. In
Sec.~\ref{fsssec} we discuss the general strategy we use to
investigate the nature of the low-temperature critical behavior.
Then, in Sec.~\ref{resultsncl3} we report the numerical results for
lattice models with $N_f=3,\,4$ and $N_c=3,\,4$.  Finally, in
Sec.~\ref{conclu} we summarize and draw our conclusions.  In
App.~\ref{largebeta} we report some results on the minimum-energy
configurations of the models considered.

\section{The multiflavor lattice SO($N_c$) gauge model}
\label{model}

We define a 2D lattice scalar gauge theory by partially gauging a
maximally symmetric model of real matrix variables $\phi_{\bm
  x}^{af}$, with $a=1,..,N_c$ and $f=1,...,N_f$ (we will refer to
these two indices as {\em color} and {\em flavor} indices,
respectively). We start from the action
\begin{eqnarray}
S_{\rm sym} = - t \sum_{{\bm x},\mu} {\rm Tr}\,\phi_{\bm
  x}^t \phi_{{\bm x}+\hat\mu} \,, \qquad {\rm Tr}\,\phi_{\bm
  x}^t \phi_{\bm x} = 1\,,
\label{ullimit}
\end{eqnarray} 
where the sum is over all links of a square lattice and
$\hat{\mu}=\hat{1},\hat{2}$ denote the unit vectors along the lattice
directions.\footnote{ Model (\ref{ullimit}) with the unit-length
  constraint for the $\phi_{\bm x}$ variables is a particular limit of
  a model with a quartic potential $\sum_{\bm x} V( {\rm Tr}\,
  \phi^\dagger_{\bm x}\,\phi_{\bm x})$ of the form $V(X) = r X +
      {1\over 2} u\, X^2$. Formally it can be obtained by setting
      $r+u=0$ and taking the limit $u\to\infty$.} Without loss of
generality, we can set $t=1$.  The action $S_{\rm sym}$ is invariant
under global O($M$) transformations with $M=N_f N_c$.  Indeed, it can
be written in terms of $M$-component unit-length real vectors ${\bm
  s}_{\bm x}$, as $S_{\rm sym} = - \sum_{{\bm x},\mu} {\bm s}_{\bm
  x}\cdot {\bm s}_{{\bm x}+\hat{\mu}}$, which is the standard
nearest-neighbor $M$-vector lattice model.

We proceed by gauging some of the degrees of freedom using the Wilson
approach~\cite{Wilson-74}. We associate an SO($N_c$) matrix $V_{{\bm
    x},\mu}$ with each lattice link [$({\bm x},\mu)$ denotes the link
  that starts at site ${\bm x}$ in the $\hat\mu$ direction] and add a
Wilson kinetic term \cite{Wilson-74} for the gauge fields.  We thus
obtain the model with action
\begin{eqnarray}
S_g  = - N_f \sum_{{\bm x},\mu} 
{\rm Tr} \,\phi_{\bm x}^t \, V_{{\bm x},\mu}
\, \phi_{{\bm x}+\hat{\mu}} 
- {\gamma\over N_c}  \sum_{{\bm x}} {\rm Tr}\,
\Pi_{\bm x}\,,\;
\label{hgauge}
\end{eqnarray}
where $\Pi_{\bm x}$ is the plaquette operator
\begin{equation}
\Pi_{\bm x}= 
V_{{\bm x},1} \,V_{{\bm x}+\hat{1},2} 
\,V_{{\bm x}+\hat{2},1}^t  
\,V_{{\bm x},2}^t
\,.
\label{plaquette}
\end{equation}
The plaquette parameter $\gamma$ plays the role of inverse gauge
coupling. The partition function reads
\begin{eqnarray}
Z = \sum_{\{\phi,V\}} e^{-\beta \,S_g}\,,\qquad \beta\equiv 1/T\,.
\label{partfun}
\end{eqnarray}
On can easily check that the lattice model (\ref{hgauge}) is invariant
under SO($N_c$) gauge transformations:
\begin{equation}
\phi_{\bm x}\to W_{\bm x} \phi_{\bm x}\,,\qquad
V_{{\bm x},\mu} \to W_{\bm x} V_{{\bm x},\mu} W_{{\bm x} +
  \hat{\mu}}^t\,,
\label{gautra}
\end{equation}  
with $W_{\bm x}\in {\rm SO}(N_c)$.  For $\gamma\to\infty$, the link
variables $V_{\bm x}$ become equal to the identity (modulo gauge
transformations), thus one recovers the ungauged model
(\ref{ullimit}), or equivalently the nearest-neighbor $M$-vector
model.

For $N_c=2$ the global symmetry group of model (\ref{hgauge}) is actually
larger than O($N_f$).  Indeed, one can show that~\cite{BPV-20-on3} 
the model can be exactly mapped onto the lattice Abelian-Higgs model
\begin{eqnarray}
&&S_{\rm AH} = - N_f \,\sum_{{\bm x}, \mu} {\rm Re} \,
  [\bar{\bm{z}}_{\bm x} \cdot \lambda_{{\bm x},\mu}\, {\bm z}_{{\bm
        x}+\hat\mu}]
\label{gllf}\\
&&\qquad -\gamma \sum_{{\bm x},\mu>\nu} 
{\rm Re}\,[
\lambda_{{\bm x},{\mu}} \,\lambda_{{\bm x}+\hat{\mu},{\nu}} 
\,\bar{\lambda}_{{\bm x}+\hat{\nu},{\mu}}  
  \,\bar{\lambda}_{{\bm x},{\nu}} ]\,,
\nonumber
\end{eqnarray}
where ${\bm z}_{\bm x}$ is a unit-length $N_f$-component complex
vector, and $\lambda_{{\bm x},{\nu}}$ a U(1) link variable.  The
Abelian-Higgs model is invariant under local U(1) and global U$(N_f)$
transformations. There is therefore an enlargement of the global
symmetry of the model: the global symmetry group is U$(N_f)$ instead
of O($N_f$). The asymptotic zero-temperature behavior of these models
have been studied in Ref.~\cite{BPV-19-ah2}. Therefore, in the
following we focus on the asymptotic low-temperature behavior for $N_c
\ge 3$.

We mention that the phase diagram and critical behavior of model
(\ref{hgauge}) in three dimensions was already discussed in
Refs.~\cite{BPV-20-on3,PV-19}, and similar results were presented in
Refs.~\cite{BPV-19-qcd3,BPV-20-qcd3} for SU($N_c$) gauge theories.  In
this work we focus on the two-dimensional case.  According to the
Mermin-Wagner theorem~\cite{MW-66}, lattice SO($N_c$) gauge theories
are not expected to show finite-temperature transitions with a
low-temperature phase in which the global O($N_f$) symmetry is
broken. Therefore, there are only two possibilities: either the system
is always disordered for any $\beta$ or a finite-temperature
transitions occurs with a low-temperature phase in which there is no
long-range order, but correlations decay algebraically with the
distance. We expect the first behavior whenever the global symmetry
group is nonabelian, the second one whenever the symmetry group is
isomorphic to U(1).

For $N_f \ge 3$, the global O($N_f$) symmetry group is
nonabelian. Therefore, we expect a nontrivial critical behavior only
in the zero-temperature limit, analogous to that occurring in the
nonlinear O($N$) $\sigma$ model or in the CP$^{N-1}$ model, see, e.g.,
Ref.~\cite{ZJ-book}. Infinite-volume correlation functions are
characterized by a length scale $\xi$ that diverges as
\begin{equation}
\xi \sim \beta^p e^{c\beta}\,.
\label{xidiv}
\end{equation}
For $N_f=2$ and $N_c\ge 3$, the model has an abelian O(2) global
symmetry.  It is therefore possible that it undergoes a
finite-temperature Berezinskii-Kosterlitz-Thouless
transition~\cite{KT-73,Berezinskii-70,Kosterlitz-74,JKKN-77,PV-13},
with a spin-wave low-temperature phase characterized by correlation
functions decaying algebraically.  For $N_f = 2$ and $N_c = 2$, due to
the mapping to the Abelian-Higgs model (\ref{gllf}), the global
symmetry group, the U(2) group, is nonabelian.  Therefore, the model
is only critical for $\beta\to\infty$. The low-temperature behavior
belongs to the universality class of the 2D CP$^1$
model~\cite{BPV-19-ah2}, which is equivalent to that of the nonlinear
O(3) $\sigma$ model.

The global symmetry group of the model is O($N_f$), which is not a
simple group.  Therefore, in principle, one may have both the breaking
the ${\mathbb Z}_2$ subgroup and of the SO($N_f$) subgroup. However,
on the basis of the results for the same model in three dimensions
\cite{BPV-20-on3} we do not expect the ${\mathbb Z}_2$ subgroup to
play any role (a similar decoupling occurs in the unitary case
\cite{BPV-20-qcd2}).  The critical low-temperature behavior is
therefore associated with the order parameter for the breaking of the
SO($N_f$) subgroup, which is the bilinear operator
\begin{equation}
Q_{\bm x}^{fg} = \sum_a \phi_{\bm x}^{af} \phi_{\bm x}^{ag} - 
    {1\over N_f} \delta^{fg}\,,
\label{qdef}
\end{equation}
which is a symmetric and traceless $N_f\times N_f$ matrix.

In the following sections we provide numerical evidence that, for
$N_c\ge 3$ and $N_f\ge 3$, the asymptotic zero-temperature limit of
the SO($N_c$) gauge model (\ref{hgauge}) is the same as that of the 2D
RP$^{N_f-1}$ models, which are also invariant under O($N_f$)
transformations. The RP$^{N-1}$ models are defined by associating a
real $N$-component unit-length vector $\varphi_{\bm x}$ with each
lattice site and considering actions that are invariant under global
O($N$) rotations of the fields and local ${\mathbb Z}_2$
transformations $\varphi_{\bm x} \to s_x \varphi_{\bm x}$ ($s_x = \pm
1$).  The standard nearest-neighbor RP$^{N-1}$ model is defined by the
lattice action
\begin{eqnarray}
S_{\rm RP} = - t \sum_{{\bm x},\mu} (\varphi_{\bm
  x} \cdot \varphi_{{\bm x}+\hat\mu})^2 \,. 
\label{rpnmodel}
\end{eqnarray} 
Alternatively, one may introduce an explicit link variable
$\sigma_{{\bm x},\mu}=\pm 1$, and consider the lattice action
\begin{eqnarray}
S_{\rm RP\sigma} = - t \sum_{{\bm x},\mu} 
    \varphi_{\bm x} \cdot \sigma_{{\bm
    x},\mu} 
\,\varphi_{{\bm x}+\hat\mu} \,.
\label{rpnmodeleps}
\end{eqnarray} 
The nature of their low-temperature behavior for $N\ge 3$ has been the
object of a long debate, see, e.g.,
Refs.~\cite{CEPS-93,Hasenbusch-96,NWS-96,CHHR-98,BFPV-20}.  The main
question has been whether the 2D RP$^{N-1}$ model belongs to the same
universality class as the O($N$) vector model.  We refer to
Ref.~\cite{BFPV-20} for a thorough discussion of this point. There, we
report extensive numerical results that indicate that the
long-distance universal behavior of the 2D RP$^{N-1}$ models differs
from that of the 2D O($N$) vector models: In the low-temperature limit
they appear as distinct universality classes.

In this work we will show that renormalization-group invariant
quantities defined in terms of $Q^{fg}$ in the nonabelian gauge theory
have the same universal behavior as the corresponding RP$^{N_f-1}$
quantities defined in terms of the local gauge-invariant operator
\begin{equation}
P_{\bm x}^{fg} = \varphi_{\bm x}^{f} \, \varphi_{\bm x}^{g}  
     - {1\over N_f} \delta^{fg} \,.
\label{prpn}
\end{equation}
Such correspondence can be established using the same arguments we
used for unitary models in Ref.~\cite{BPV-20-qcd2}. As discussed in
the Appendix, for $\beta\to\infty$ the $\phi$ configurations can be
parametrized by a single $N_f$-dimensional unit vector
$\varphi^f$. Modulo gauge transformations, we have
\begin{eqnarray}
     \phi^{af} = 0\quad  && \qquad \hbox{$a < N_c$} \nonumber \\
     \phi^{af} = \varphi^f  && \qquad \hbox{$a =  N_c$}
\end{eqnarray}
which implies that the bilinear $Q_{\bm x}$ becomes equivalent in this
limit to the RP$^{N_f-1}$ operator $P_{\bm x}$. Since the ${\mathbb
  Z}_2$ global symmetry does not play any role, in the zero
temperature-limit the gauge model can be described by an effective
theory only in terms of the SO($N_f$) order parameter $P_{\bm x}$.
The natural candidate for the action is
\begin{equation}
H_{\rm eff} = - \kappa \sum_{{\bm x},\mu}\hbox{Tr}\, P_{\bm x} P_{{\bm x}
  +\hat{\mu}}\,,
\end{equation}
which gives (\ref{rpnmodel}) apart from an irrelevant constant. We
have thus obtained the RP$^{N_f-1}$ model.

\section{Universal finite-size scaling}
\label{fsssec}

We exploit FSS techniques~\cite{FB-72,Barber-83,Privman-90,PV-02} to
study the nature of the asymptotic critical behavior of the model for
$T\to 0$.  For this purpose we consider models defined on square
lattices of linear size $L$ with periodic boundary conditions.  We
focus on the correlations of the gauge-invariant variable $Q_{\bm x}$
defined in Eq.~(\ref{qdef}). The corresponding two-point correlation
function is defined as
\begin{equation}
G({\bm x}-{\bm y}) = \langle {\rm Tr}\, Q_{\bm x} Q_{\bm y} \rangle\,,  
\label{gxyp}
\end{equation}
where the translation invariance of the system has been taken into
account. We define the susceptibility $\chi=\sum_{\bm x} G({\bm x})$
and the correlation length
\begin{eqnarray}
\xi^2 = {1\over 4 \sin^2 (\pi/L)}
{\widetilde{G}({\bm 0}) - \widetilde{G}({\bm p}_m)\over 
\widetilde{G}({\bm p}_m)}\,,
\label{xidefpb}
\end{eqnarray}
where $\widetilde{G}({\bm p})=\sum_{{\bm x}} e^{i{\bm p}\cdot {\bm x}}
G({\bm x})$ is the Fourier transform of $G({\bm x})$, and ${\bm p}_m =
(2\pi/L,0)$. We also consider the quartic cumulant (Binder) parameter
defined as
\begin{equation}
U = {\langle \mu_2^2\rangle \over \langle \mu_2 \rangle^2} \,, \qquad
\mu_2 = {1\over V^2}
\sum_{{\bm x},{\bm y}} {\rm Tr}\,Q_{\bm x} Q_{\bm y}\,,
\label{binderdef}
\end{equation}
where $V=L^2$.

To identify the universality class of the asymptotic zero-temperature
behavior, we consider the Binder parameter $U$ as a function of the
ratio
\begin{equation}
R_\xi\equiv \xi/L\,.
\label{rxidef}
\end{equation}
Indeed, in the FSS limit we have (see, e.g., Refs.~\cite{BPV-19-ah2})
\begin{equation}
U(\beta,L) \approx F(R_\xi)\,,
\label{r12sca}
\end{equation}
where $F(x)$ is a universal scaling function that completely
characterizes the universality class of the transition.  The
asymptotic values of $F(R_{\xi})$ for $R_{\xi}\to 0$ and
$R_{\xi}\to\infty$ correspond to the values that $U$ takes in the
small-$\beta$ and large-$\beta$ limits.  For $R_\xi\to 0$ we have
\begin{eqnarray}
\lim_{R_\xi\to 0} \;U = 1 + {4\over (N_f-1) (N_f+2)}\, .
\label{uextr}
\end{eqnarray}
independently of the value of $N_c$. In the large-$\beta$ limit we 
have  $U\to 1$ as discussed in App.~\ref{largebeta}.

Eq.~(\ref{r12sca}) allows us to check the universality of the
asymptotic zero-temperature behavior without the need of tuning any
parameter.  Corrections to Eq.~(\ref{r12sca}) decay as a power of
$L$. In the case of asymptotically free models, such as the 2D
CP$^{N-1}$ and O($N$) vector models, corrections decrease as $L^{-2}$,
multiplied by powers of $\ln L$ ~\cite{CP-98}.  However we note that
sometimes, when the available data are not sufficiently asymptotic, the
approach to the asymptotic behavior may appear slower, and corrections
apparently decay as $L^{-p}$ with $p < 2$ \cite{BNW-10}. 

Because of the universality of relation (\ref{r12sca}), we can use the
plots of $U$ versus $R_\xi$ to identify the models that belong to the
same universality class. If the data of $U$ for two different models
follow the same curve when plotted versus $R_\xi$, their critical
behavior is described by the same continuum quantum field theory.
This implies that any other dimensionless RG invariant quantity has
the same critical behavior in the two models, both in the
thermodynamic and in the FSS limit.  An analogous strategy for the
study of the asymptotic zero-temperature behavior of 2D models was
employed in Refs.~\cite{BPV-19-ah2,BPV-20-qcd2}.

\section{Numerical results}
\label{resultsncl3}

\begin{figure}[tbp]  
\includegraphics[width=0.95\columnwidth]{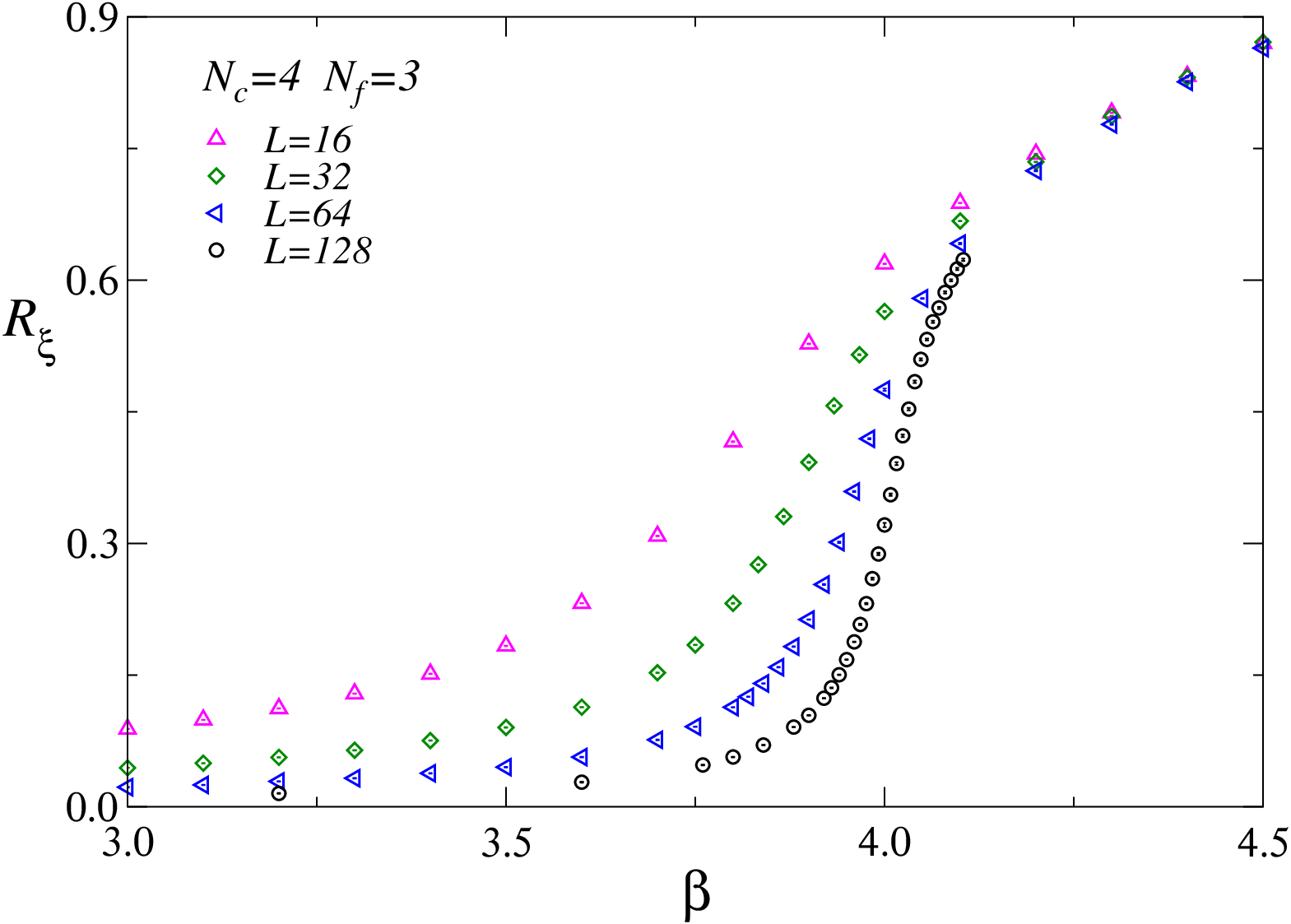}
\includegraphics[width=0.95\columnwidth]{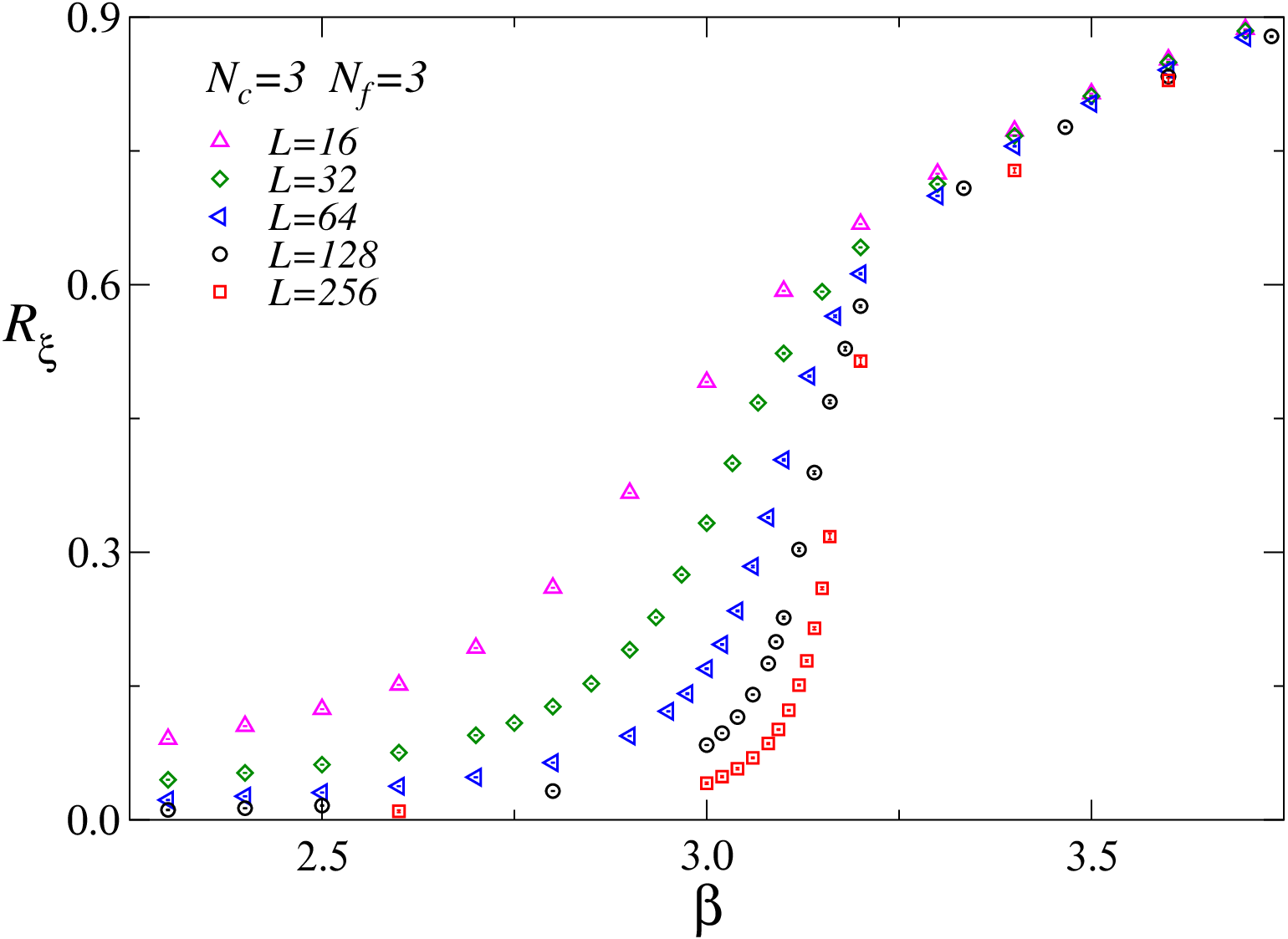}
\caption{$R_\xi\equiv \xi/L$ for the three-flavor SO(3) and SO(4)
  gauge theories (\ref{hgauge}) with $\gamma=0$.  We show data up to
  $L=256$ for $N_c=3$ (bottom) and up $L=128$ for $N_c=4$ (top).  Data
  for different sizes do not show evidence of crossing points.  
  Statistical errors are hardly visible on the scale of the figure.
  }
\label{Rxinf3nc34}
\end{figure}

\begin{figure}[tbp]  
\includegraphics[width=0.95\columnwidth]{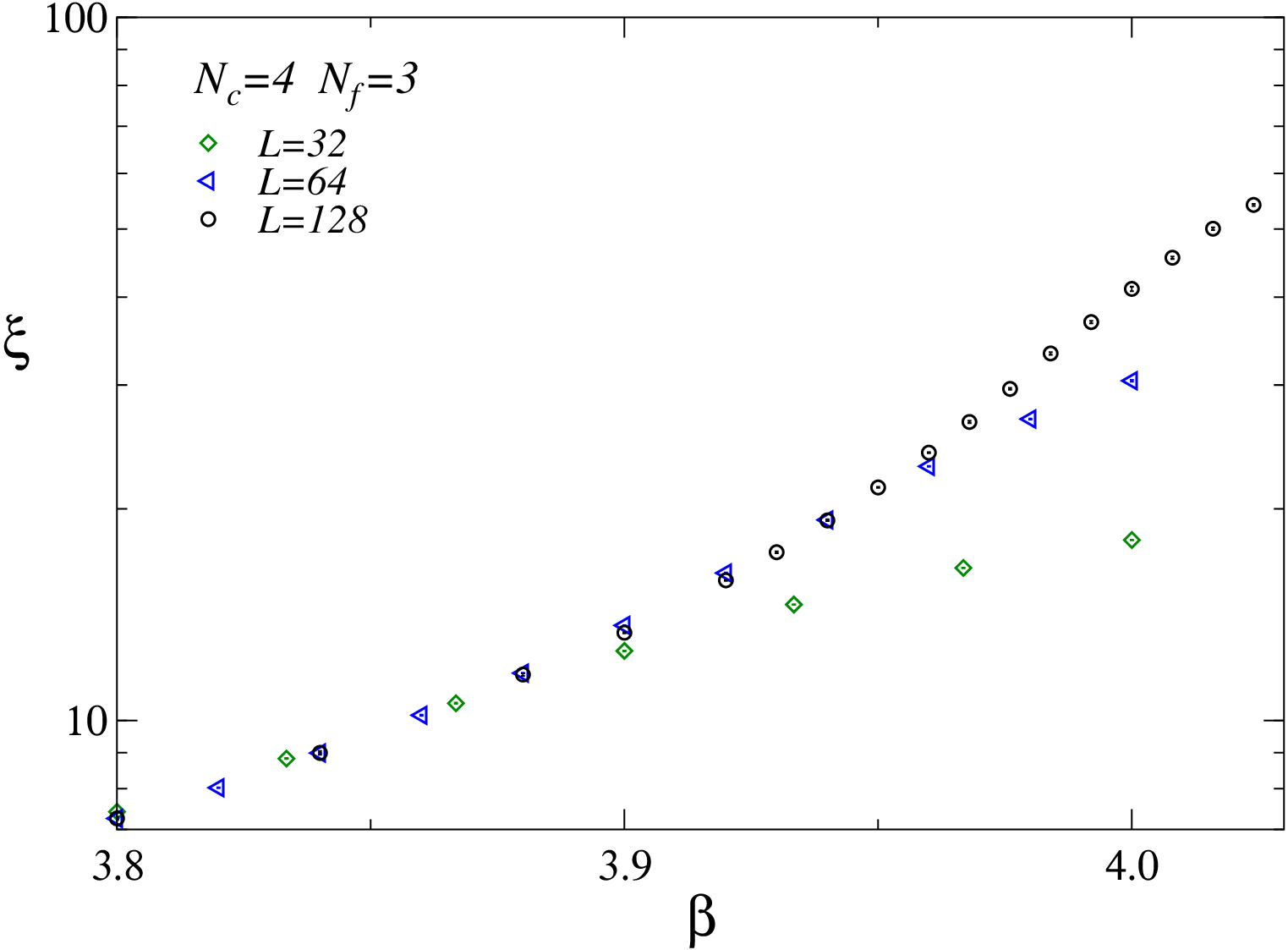}
\includegraphics[width=0.95\columnwidth]{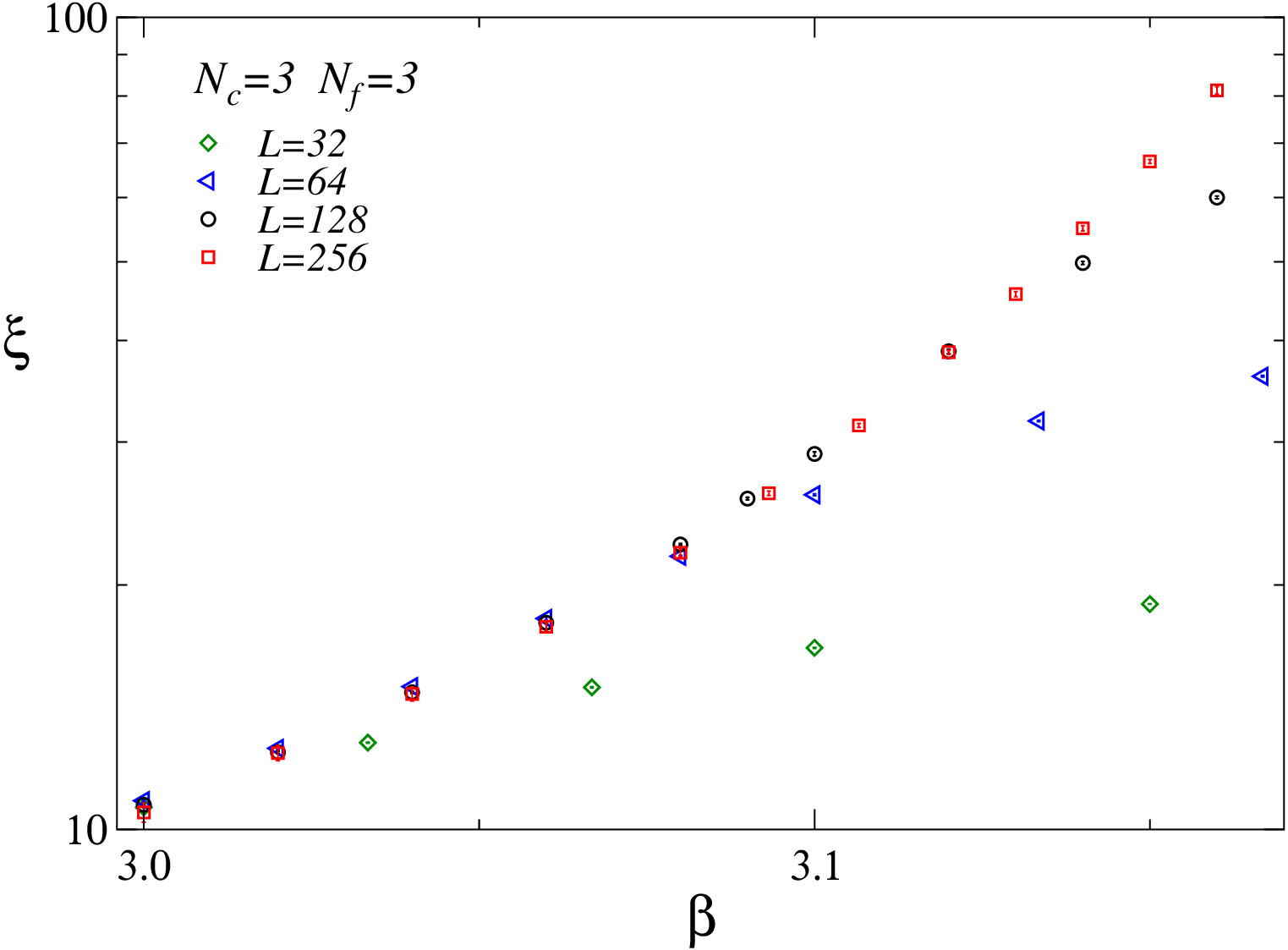}
\caption{The correlation length $\xi$ versus $\beta$ for $N_f=3$,
  $N_c=3$ (bottom) and $N_f=3$ $N_c=4$ (top). We set $\gamma=0$.  When
  data for different values of $L$ match, they may be considered as
  good approximations of the infinite-volume correlation length,
  within their errors.  The behavior of the infinite-volume data is
  consistent with an exponential dependence on $\beta$ (we use a
  logarithmic scale on the vertical axis).  }
\label{xinf3nc34}
\end{figure}

\begin{figure}[tbp]  
\includegraphics[width=0.95\columnwidth]{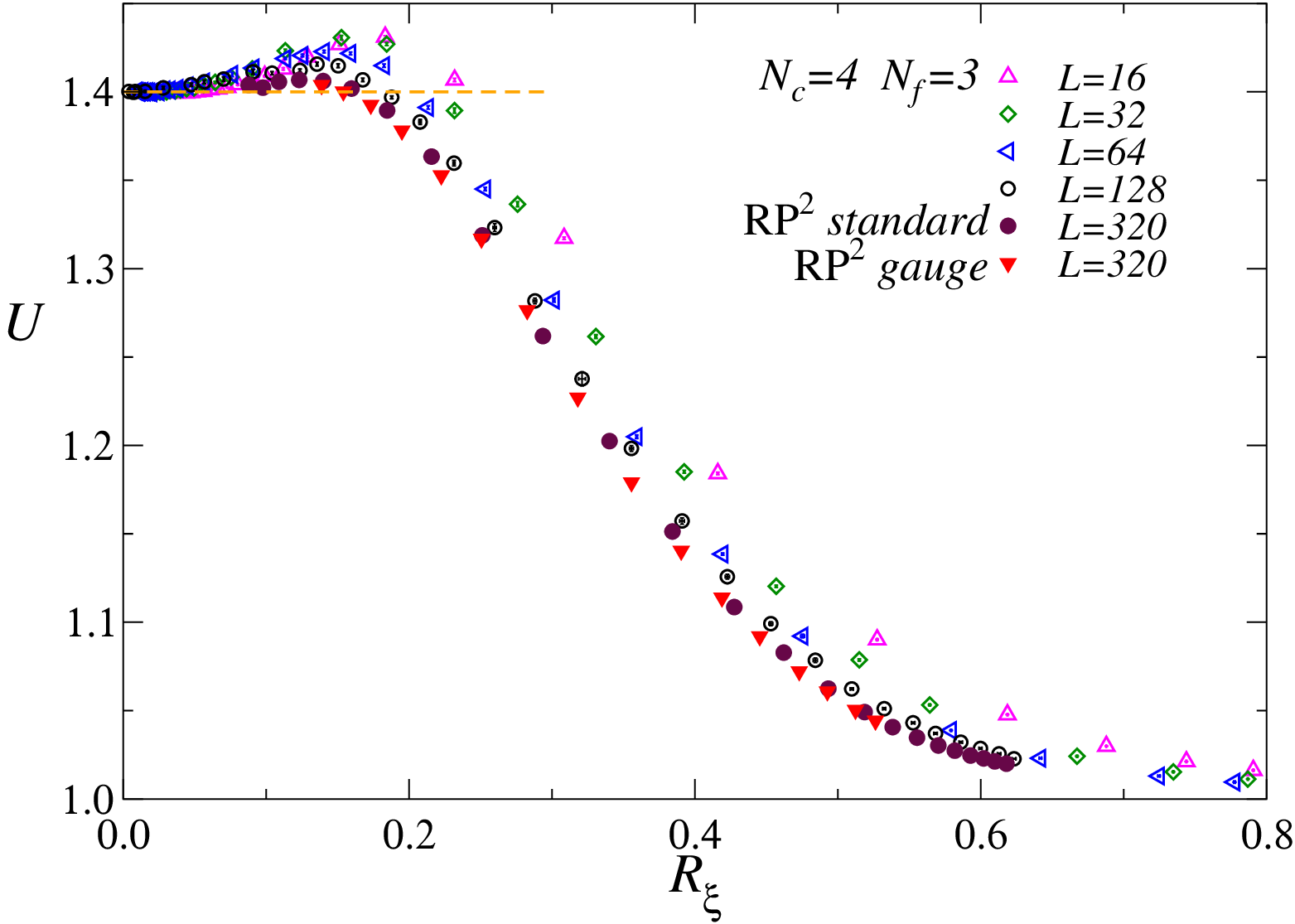}
\includegraphics[width=0.95\columnwidth]{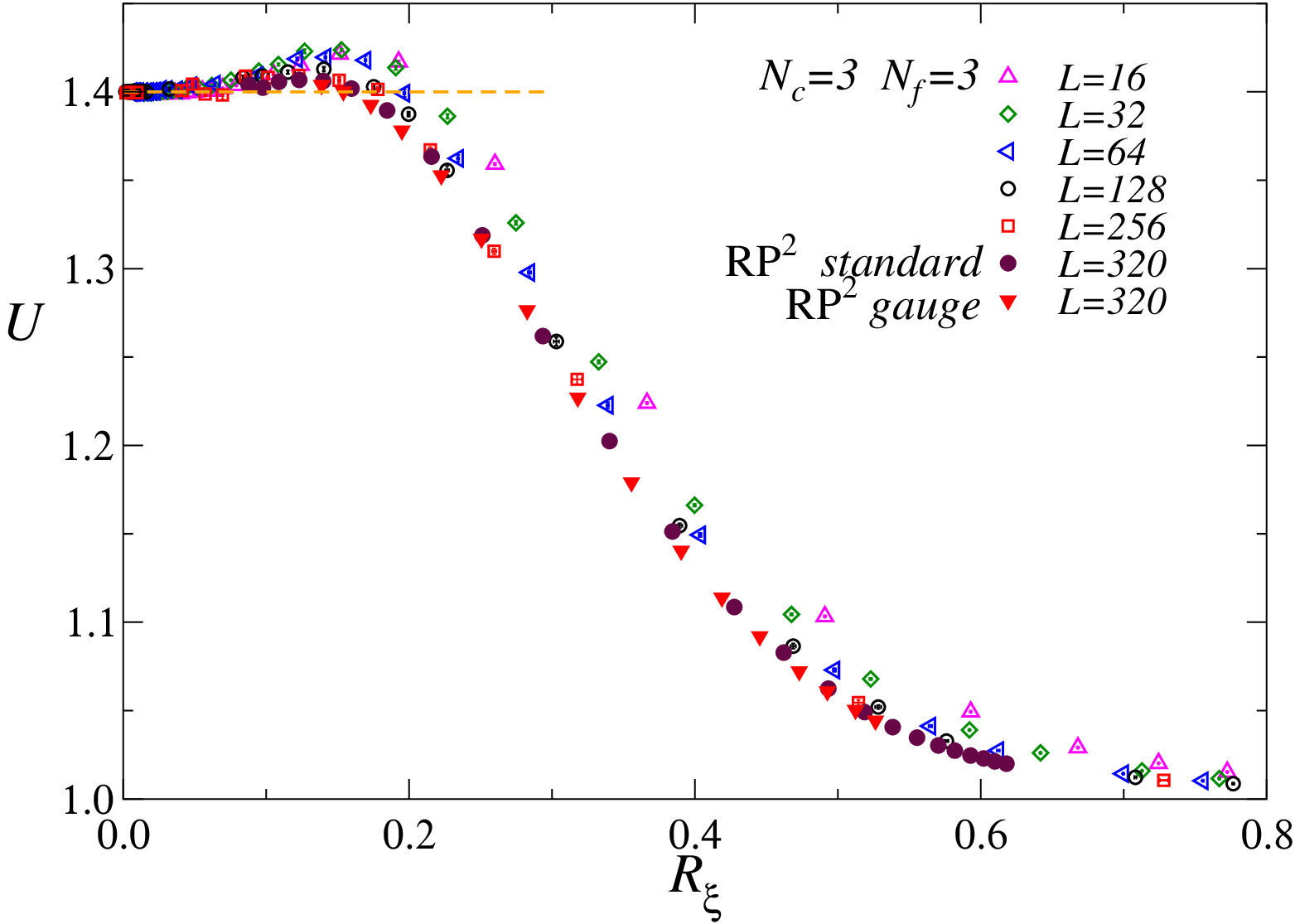}
\caption{Plot of $U$ versus $R_\xi$ for the three-flavor SO(3)
  (bottom) and SO(4) (top) gauge theory at $\gamma=0$.  The data
  approach the asymptotic curve of the 2D RP$^2$ models
  (\ref{rpnmodel}) and (\ref{rpnmodeleps}) (labelled as {\em
    standard} and {\em gauge}, respectively; the corresponding data for
  $L=320$ are taken from Ref.~\cite{BFPV-20}).
  Statistical errors are so small to be hardly visible.}
\label{u-rxi-nf3nc34}
\end{figure}

\begin{figure}[tbp]
\includegraphics[width=0.95\columnwidth]{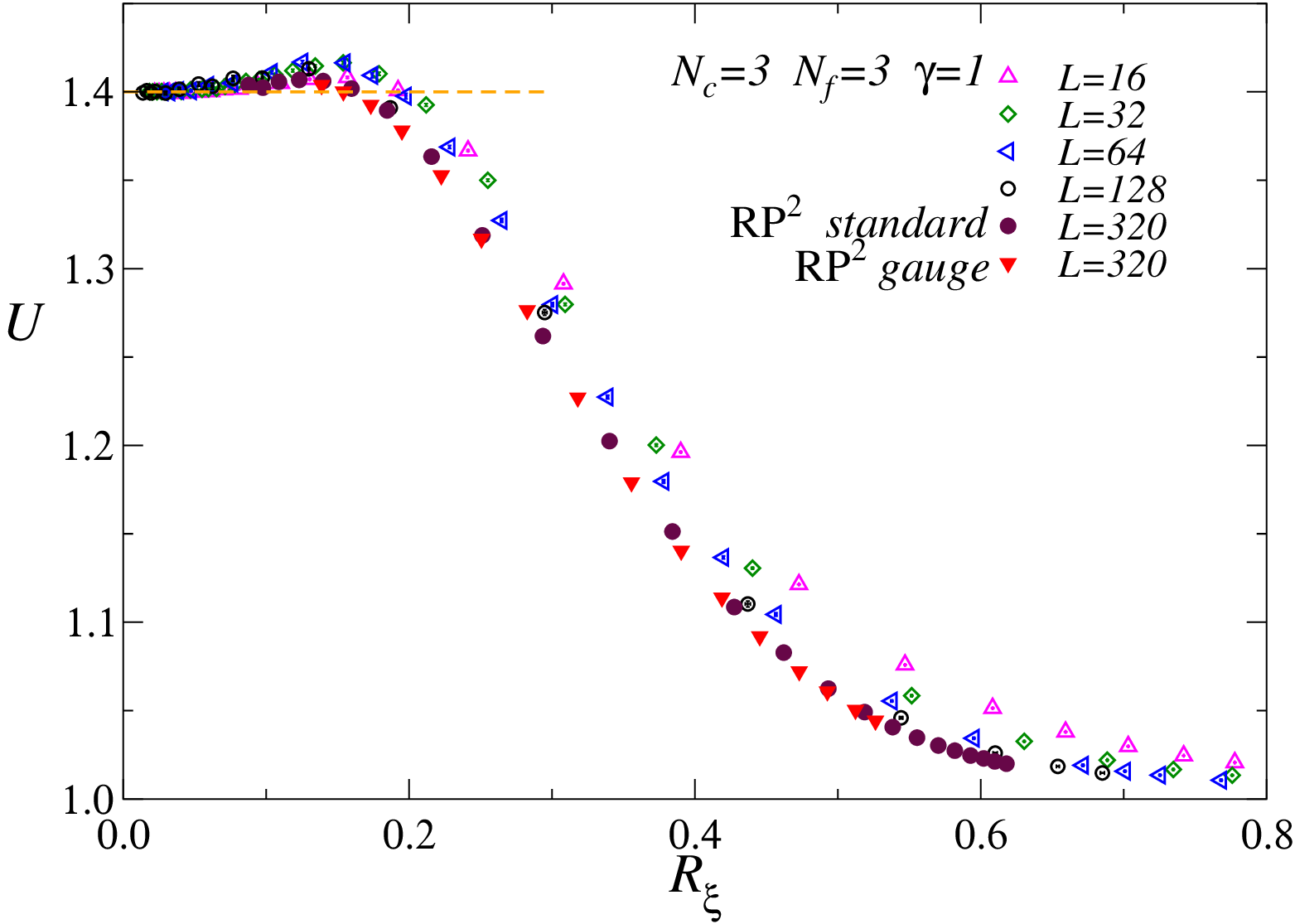}
\includegraphics[width=0.95\columnwidth]{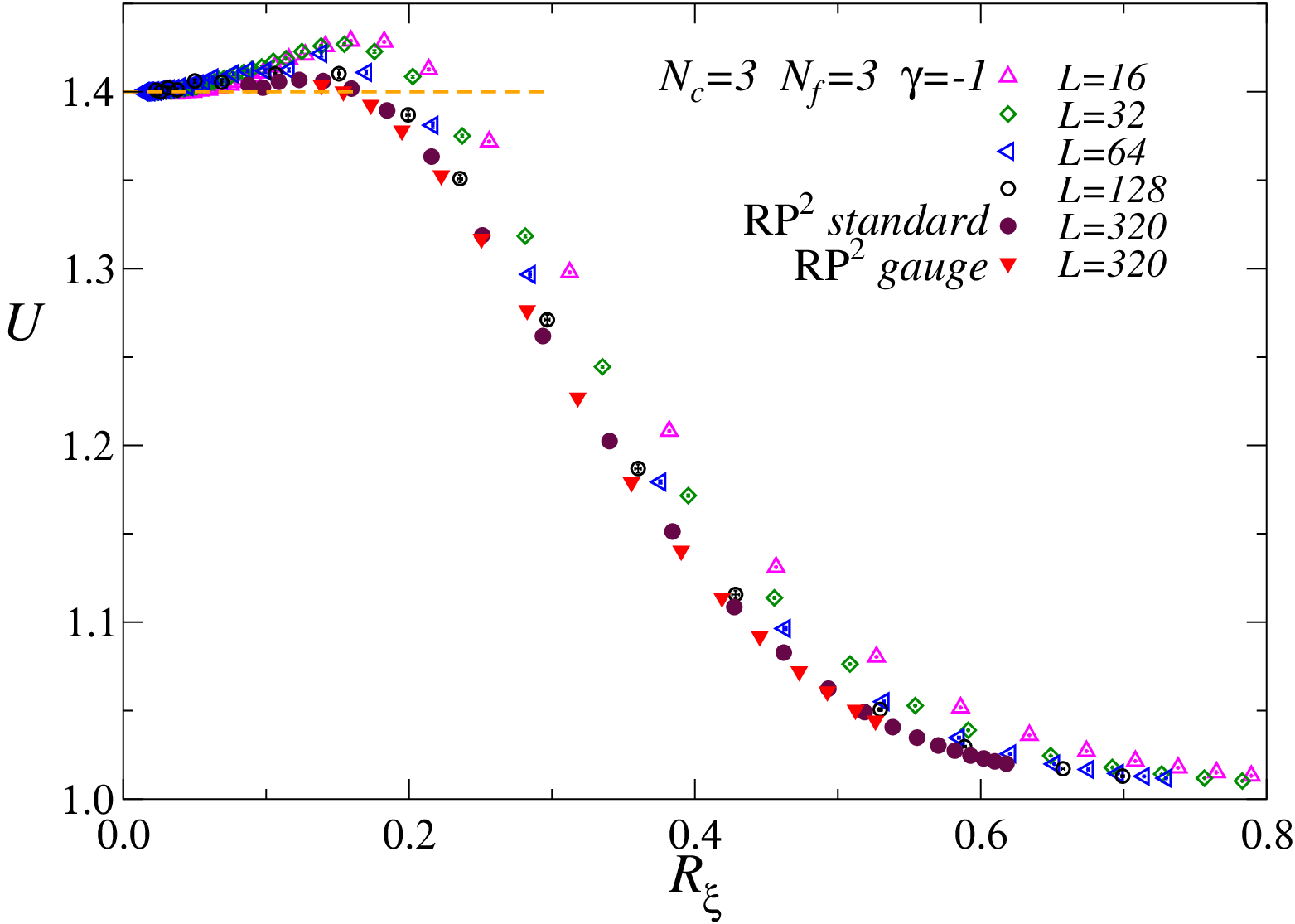}
\caption{Plot of $U$ versus $R_\xi$ for $N_f=3$, $N_c=3$, $\gamma=-1$
  (lower panel) and $\gamma=1$ (upper panel). Data approach the same
  universal FSS curve obtained for the $\gamma=0$ SO($N_c$) gauge
  model and the RP$^2$ models (\ref{rpnmodel}) and (\ref{rpnmodeleps})
  (see Fig.~\ref{u-rxi-nf3nc34}).  }
\label{u-rxi-nf3nc3gamma}
\end{figure}

\begin{figure}[tbp]
\includegraphics[width=0.95\columnwidth]{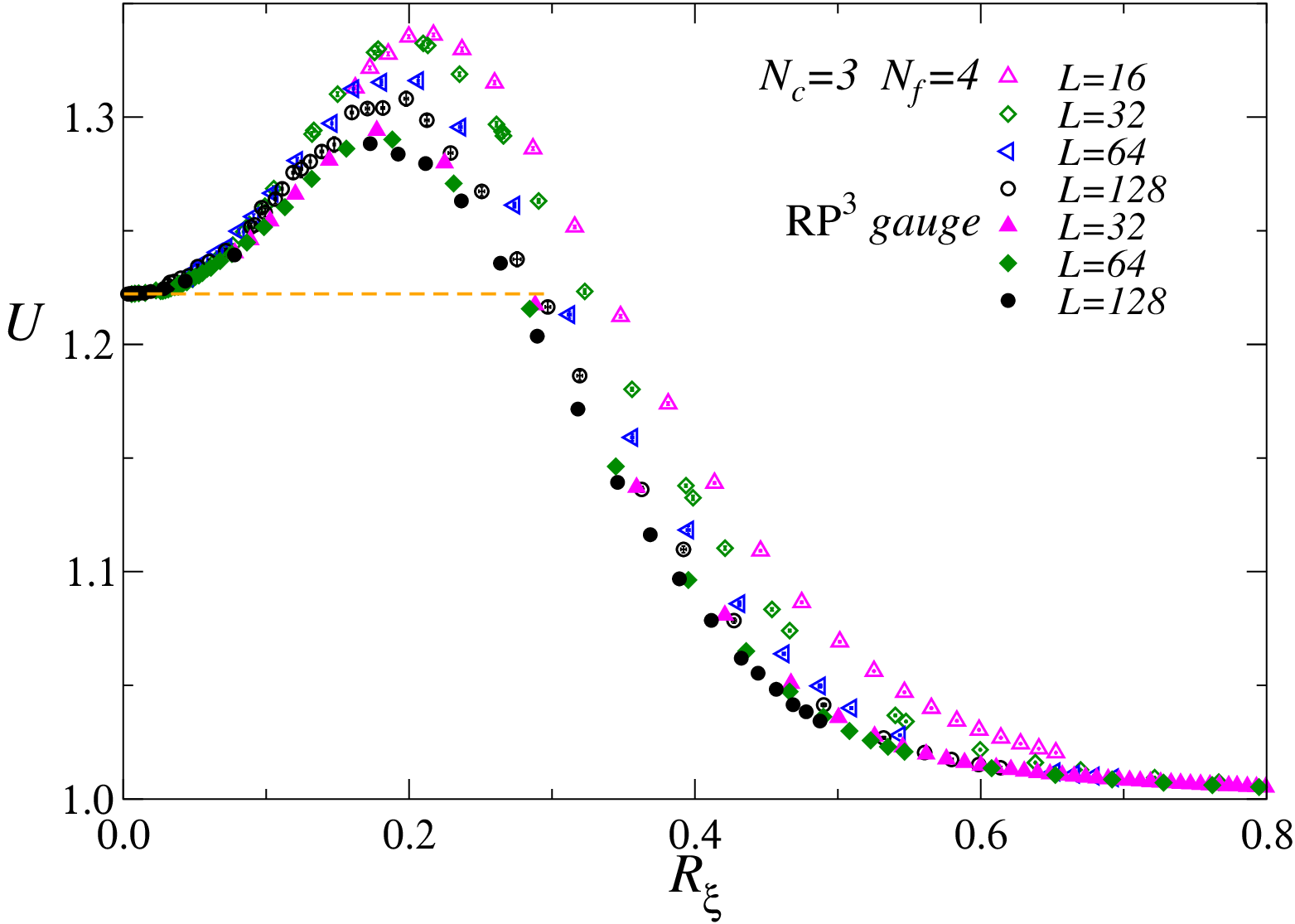}
\caption{Plot of $U$ versus $R_\xi$ for $N_f=4$, $N_c=3$, and
  $\gamma=0$.  Data approach the same universal FSS curve obtained for
  the RP$^3$ models (\ref{rpnmodel}) and (\ref{rpnmodeleps}) . }
\label{u-rxi-nf4nc3}
\end{figure}

In this section we study the large-$\beta$ critical behavior of the
lattice scalar gauge model (\ref{hgauge}) for some values of $N_f\ge
3$ and $N_c\ge 3$.  We perform MC simulations, using the same
upgrading algorithm employed in three dimensions \cite{BPV-20-on3}.
We show that the FSS curves (\ref{r12sca}) of the Binder parameter $U$
versus $R_\xi$ computed in the model (\ref{hgauge}) agree with those
computed in RP$^{N-1}$ models (we use the results reported in
Ref.~\cite{BFPV-20}). These results provide numerical evidence that,
for $N_c\ge 3$, the critical behavior belongs to the universality
class of the 2D RP$^{N_f-1}$ field theory, in agreement with the
arguments of the previous section.

We first mention that the data of $R_\xi\equiv \xi/L$ corresponding to
different lattice sizes, see Fig.~\ref{Rxinf3nc34}, do not intersect,
confirming the absence of a phase transition at finite $\beta$, as
expected on the basis of the Mermin-Wagner theorem~\cite{MW-66}.  In
Fig.~\ref{xinf3nc34}, we show the estimates of the correlation length
for the three-flavor SO(3) and SO(4) gauge theories (\ref{hgauge})
with $\gamma=0$, up to lattice sizes $L=256$ and $L=128$,
respectively.  When data for different lattice sizes match, they can
be considered as a good approximation of the correlation length in
thermodynamic limit at the given inverse temperature $\beta$. The data
in this regime are substantially consistent with an exponential
dependence of $\xi$ on $\beta$, see Eq.~(\ref{xidiv}), as expected for
asymptotically free models.

In Fig.~\ref{u-rxi-nf3nc34} we plot $U$ versus $R_\xi$ for the
three-flavor SO(3) and SO(4) gauge theories with $\gamma=0$, up to
$L=256$ and $L=128$, respectively. We observe that the data of $U$
appear to approach a FSS curve in the large-$L$ limit, in agreement
with the FSS prediction (\ref{r12sca}).  In the same figure we also
report data for the standard RP$^2$ lattice model with
action~(\ref{rpnmodel}), and for the RP$^2$ gauge model with action
~(\ref{rpnmodeleps}) (as shown in Ref.~\cite{BFPV-20}, the data for
$L=320$ provide a good approximation of the asymptotic curve). The
RP$^2$ results are consistent with the asymptotic FSS curve for the
SO($N_c$) gauge model, confirming our claim that the RP$^2$ model and
the SO($N_c$) gauge model with $N_f=3$ and any $N_c\ge 3$ have the
same large-distance universal behavior in the critical limit
$\beta\to\infty$.

We have also performed MC simulations for nonvanishing values of
$\gamma$.  Fig.~\ref{u-rxi-nf3nc3gamma} reports data for the
three-flavor SO(3) gauge theory (\ref{hgauge}) with $\gamma=\pm 1$, up
to $L=128$.  They appear to approach the asymptotic FSS curve of the
RP$^2$ universality class, demonstrating that the universal features
of the asymptotic low-temperature behavior are independent of the
inverse gauge coupling $\gamma$, at least in a wide interval around
$\gamma=0$. Data up to $L=64$ for $\gamma=\pm 2$ (not shown) also
approach the RP$^2$ curve as $L$ increases.  As discussed in
Sec.~\ref{model}, the asymptotic FSS curves must change if we take the
limit $\gamma\to\infty$ and then the limit $\beta\to\infty$, In this
case the SO(3) and SO(4) gauge theories turn into the O(9) and O(12)
model, respectively.

These results should be considered as a robust evidence that the
asymptotic low-temperature behavior of the three-flavor lattice gauge
theory with SO(3) and SO(4) gauge symmetry belongs to the universality
class of the 2D RP$^2$ universality class, in a large interval of
values of $\gamma$ around $\gamma=0$.

As an additional check of the arguments presented in Sec.~\ref{model},
we have performed simulations of the model (\ref{hgauge}) for $N_f=4$,
$N_c=3$ and $\gamma=0$. The results for the Binder parameter are
plotted versus $ R_\xi$ in Fig.~\ref{u-rxi-nf4nc3}.  For comparison we
also report results for the RP$^3$ gauge model.  The SO(3) gauge data
show a significant size dependence, but with a clear trend towards the
RP$^3$ data. In particular, the SO(3) gauge data corresponding to
$L=128$ are essentially consistent with the RP$^3$ data, confirming
again the asymptotic equivalence of the universal large-distance
behavior of the SO(3) gauge model and of the RP$^3$ model.

\section{Conclusions}
\label{conclu}

We have studied a class of 2D lattice nonabelian SO($N_c$) gauge
models with multicomponent scalar fields, focusing on the role that
global and local nonabelian gauge symmetries play in determining the
universal features of the asymptotic low-temperature behavior.  Such
lattice gauge models are obtained by partially gauging a maximally
O($M$)-symmetric multicomponent scalar model, $M=N_fN_c$, 
using the Wilson lattice approach. For $N_c \ge 3$, 
the resulting theory is locally invariant under SO($N_c$)
gauge transformations and globally invariant under O($N_f$)
transformations. For $N_c=2$, these 
lattice gauge models are instead equivalent
to the 2D Abelian-Higgs model
and therefore have a larger U($N_f$) 
global invariance group.
The fields belong to the coset
$S^{M-1}$/SO($N_c$), where $M= N_f N_c$ and $S^{M-1}$ is the
$(M-1)$-sphere in an $M$-dimensional space. 

Since for $N_c=2$ these lattice gauge models are equivalent
to the 2D Abelian-Higgs models, already studied in
Ref.~\cite{BPV-19-ah2}, we only consider
$N_c\ge 3$. Moreover, we will only consider models with $N_f \ge 3$.
In this 
case the global symmetry group is nonabelian, and thus one expects 
the system to develop a critical behavior only in the zero-temperature limit.
For $N_f=2$ the behavior is expected to be different, since 
the global abelian O(2)
symmetry may allow finite-temperature Berezinskii-Kosterlitz-Thouless
transitions. 

The universal features of the zero-temperature behavior 
are determined by means of MC simulations.
We consider the 
lattice SO($N_c$) gauge models (\ref{hgauge}) for $N_c=3$, 4  and for 
$N_f=3,4$ The FSS analyses of the MC results
provide numerical evidence that the asymptotic
low-temperature behavior is the same as that of the 2D
RP$^{N_f-1}$ models, characterized by the same global O($N_f$)
symmetry and by a local ${\mathbb Z}_2$ gauge symmetry. 
The numerical results are supported by theoretical arguments that show
that RP$^{N_f-1}$ models and SO($N_c$) gauge theories with $N_f$ 
flavors have the same ground-state (zero-temperature) properties. 
Moreover, the gauge degrees 
of freedom decouple as $\beta\to\infty$. 

These results provide further support to the conjecture put forward in
Ref.~\cite{BPV-20-qcd2}, that the renormalization-group flow
determining the asymptotic low-temperature behavior is generally
controlled by the 2D statistical theories associated with the
symmetric spaces that have the same global symmetry. For models with 
complex fields and U($N_f$) global invariance---for instance,
the multicomponent lattice
Abelian-Higgs model and the multiflavor lattice scalar chromodynamics
considered in Ref.~\cite{BPV-20-qcd2}---the universal behavior is 
described by the 2D CP$^{N_f-1}$ field theory. 
For the lattice SO($N_c$) gauge models with $N_c\ge3$ and $N_f\ge3$, instead,
the RP$^{N_f-1}$ field theory is the relevant one.

\bigskip

\emph{Acknowledgement}.  Numerical simulations have been performed on
the CSN4 cluster of the Scientific Computing Center at INFN-PISA.

\appendix

\section{Minimum-energy configurations}
\label{largebeta}

In this appendix we identify the minimum-energy configurations for the
action (\ref{hgauge}). The analysis is very similar to that presented 
for unitary models in Ref.~\cite{BPV-20-qcd2}. We refer the reader to this 
work for additional details.

\begin{table}[tbh]
    \centering
        \caption{Estimates of several observables on the 
          minimum-energy configurations for $\gamma=0$, for two
          lattice sizes $L=4,\,8$. They are obtained by fitting 
          large-$\beta$ numerical data (we use the same
          procedure discussed in the appendix of Ref.~\cite{BPV-20-qcd2}).
          }
    \begin{tabular}{cccccc}
\hline    
         $(N_c,N_f)$& $L$ &
         $\langle{\rm Tr}\ \Pi_{\bm x}\rangle/N_c$ & 
         $S_g/(2VN_f)$ & 
         $U$&$1 + \langle {\rm Tr}\ Q_{\bm x}^2\rangle $\\
         \hline
         \hline
$(3,3)$&4&0.3504(2)&$-$1.00000(1)&1.000000(1)&1.0000(1)\\
       &8&0.3501(1)&$-$0.99999(1)& 1.000001(1) & 0.99998(1)\\
         \hline
$(3,4)$&4&0.3600(3)&$-$1.00000(1)&1.000000(1)& 1.00000(1)\\
       &8&0.3587(1)&$-$1.00000(1) & 0.999999(1)& 0.99999(1)\\
         \hline
$(4,3)$&4&0.2564(1)&$-$0.99999(1)&0.999999(1)&1.00000(1)\\
       &8&0.2563(1)&$-$0.99999(1)&0.999999(1)&1.00000(2)\\
         \hline
$(4,4)$&4&0.2599(1)&$-$1.00000(1) &1.00000(1)&1.00001(2)\\
       &8&0.2595(1)&$-$1.00000(1)&1.00000(1)&1.00000(1)\\
         \hline
    \end{tabular}
    \label{tab:gamma0}
\end{table}

We start by considering the simplest case
$\gamma=0$. The minimum-energy configurations are those that 
satisfy the condition
\begin{eqnarray}
{\rm Tr} \; [\phi_{\bm x}^t 
V_{{\bm x},\mu} \phi_{{\bm x}+\hat\mu}] = 1\,
\label{minenea}
\end{eqnarray}
for each link. This condition is satisfied if $\phi_{{\bm
    x}+\hat{\mu}} = V_{{\bm x},\mu}^t \phi_{\bm x}$, and therefore
$Q_{\bm x} = Q_{{\bm x}+\hat{\mu}}$, thus entailing the breaking of
the global symmetry for $\beta\to\infty$.  

The previous relation implies the
consistency condition $\phi_{\bm x} = \Pi_{\bm x} \,\phi_{\bm x}$,
where $\Pi_{\bm x}$ is the plaquette operator (\ref{plaquette}).
For $N_c\ge 3$, such consistency condition has several classes of
different solutions.  The plaquette $\Pi_{\bm x}$ must satisfy
\begin{equation}
\Pi_{\bm x} = A \oplus 1 = 
\begin{pmatrix} A & 0 \\
                0 & 1 
\end{pmatrix}
\label{Pi-largebeta}
\end{equation}
where $A$ is an SO($N_c-1$) matrix, modulo a gauge transformation. The
corresponding configurations of the fields $\phi_x$ depend on the
structure of the matrix $A$. If $A$ is a generic unitary matrix which
does not have unit eigenvalues, the field $\phi$ is necessarily given
by
\begin{equation}
\begin{array}{ll}
 \phi^{af} = 0 \qquad & {a < N_c}\, ,  \\
 \phi^{af} = v^f \qquad & {a = N_c}\, ,
\end{array}
\label{phi-largebeta}
\end{equation}
where $v^f$ is a unit $N_f$-dimensional vector.  Different $\phi$
configurations are only possible if $A$ has some unit eigenvalues. For
instance, if $A = A_1 \oplus 1$, with $A_1$ belonging to the
SO($N_c-2$) subgroup, then the $\phi$ field configurations of the form
\begin{equation}
\begin{array}{ll}
 \phi^{af} = 0 \qquad & {a < N_c-1}\, , \\ \phi^{af} = w^f
 \qquad & {a = N_c-1}\, , \\ \phi^{af} = v^f \qquad & {a = N_c}\, ,
\end{array}
\label{phi-lowtemp}
\end{equation}
($v^f$ and $w^f$ are generic $N_f$-dimensional vectors) satisfy the
condition $\phi_{\bm x} = \Pi_{\bm x} \,\phi_{\bm x}$.  To understand
which type of configurations dominate, we have again resorted to
numerical simulations on small lattices.  The results are reported in
Table~\ref{tab:gamma0}. For the plaquette operator $\Pi_{\bm x}$,
see Eq.~(\ref{plaquette}), results are consistent with
\begin{equation}
   \langle \hbox{Tr } \Pi_{\bm x} \rangle = 1\,,
   \label{plaquexplb}
\end{equation}
in the large-$L$ limit.  This relation is consistent with
Eq.~(\ref{Pi-largebeta}) only if we assume that the matrix $A$ is a
randomly chosen SO($N_c-1$) matrix. For instance, if $A= A_1 \oplus 1$
with a generic $A_1\in$ SO($N_c-2$), one would instead predict $\langle
\hbox{Tr } \Pi_{\bm x} \rangle = 2$.  This result constraints the
field $\phi$ to be of the form (\ref{phi-largebeta}).  If this is the
case, the operator $Q_{\bm x}$, defined in Eq.~(\ref{qdef}), takes the form
$Q_{\bm x}^{fg} = v^f v^g - \delta^{fg}/N_f$ in the large-$\beta$ 
regime.   Therefore,
$Q_{\bm x}$ becomes equivalent to the operator $P_{\bm x}$ defined in 
the RP$^{N_f-1}$ theory.
As an additional check that the 
relevant configurations are those of the form (\ref{phi-largebeta}),
we compute the Binder parameter, which should converge to
1.  The numerical results reported in Table~\ref{tab:gamma0} are in
good agreement.

When $\gamma\neq 0$ the analysis of the minimum-energy configurations
becomes more complicated, as is also the case for lattice SU($N_c$)
gauge theories (see the 
appendix of Ref.~\cite{BPV-20-qcd2}).  We do not repeat here the arguments of
Ref.~\cite{BPV-20-qcd2}. They apply to SO($N_c$) theories as well, as we 
have explicitly verified numerically for $\gamma=-1$ and $\gamma=1$.  We only
mention that, as in the case of SU($N_c$) gauge
theories, the gauge parameter $\gamma$ is relevant for 
gauge properties, but not for the behavior of the $\phi$
correlations, which dominate the large-$\beta$ limit.


\begin{thebibliography}{99}

\bibitem{Wilson-74} K.G. Wilson, Confinement of quarks, Phys. Rev. D
  {\bf 10}, 2445 (1974).

\bibitem{Weinberg-book} S. Weinberg, {\em The Quantum Theory of
  Fields}, (Cambridge University Press, 2005).

\bibitem{Sachdev-19} S. Sachdev, Topological order, emergent gauge
  fields, and Fermi surface reconstruction, Rep. Prog. Phys. {\bf 82},
  014001 (2019).

\bibitem{Anderson-book} P.~W.~Anderson, {\em Basic Notions of
  Condensed Matter Physics}, (The Benjamin/Cummings Publishing
  Company, Menlo Park, California, 1984).

\bibitem{ZJ-book} J. Zinn-Justin, 
  {\em Quantum Field Theory and Critical Phenomena}, 
  fourth edition (Clarendon Press, Oxford, 2002).

\bibitem{BPV-19-ah2} C. Bonati, A. Pelissetto and E. Vicari,
  Two-dimensional multicomponent Abelian-Higgs lattice models,
  Phys. Rev. D {\bf 101}, 034511 (2020).

\bibitem{BPV-20-qcd2} C. Bonati, A. Pelissetto, and E. Vicari,
  Universal low-temperature behavior of two-dimensional lattice scalar
  chromodynamics, Phys. Rev. D {\bf 101}, 054503 (2020).

\bibitem{BHZ-80} E. Br\'ezin, S. Hikami, and J. Zinn-Justin,
  Generalized non-linear $\sigma$-models with gauge invariance,
  Nucl. Phys. B {\bf 165}, 528 (1980).

\bibitem{MW-66} N.~D.~Mermin and H.~Wagner, Absence of ferromagnetism
  or antiferromagnetism in one- or two-dimensional isotropic
  Heisenberg models, Phys. Rev. Lett.  {\bf 17}, 1133 (1966).



\bibitem{KT-73} 
J. M. Kosterlitz and  D. J. Thouless,
  Ordering, metastability and phase transitions in two-dimensional systems,
  J.\ Phys. C: Solid State {\bf 6},  1181 (1973).

\bibitem{Berezinskii-70} V. L. Berezinskii, Destruction of Long-range
  Order in One- dimensional and Two-dimensional Systems having a
  Continuous Symmetry Group I. Classical Systems,
  Zh. Eksp. Theor. Fiz. {\bf 59}, 907 (1970) [Sov. Phys. JETP {\bf
      32}, 493 (1971)].

\bibitem{Kosterlitz-74}
  J. M. Kosterlitz, 
  The critical properties of the two- dimensional xy model,
  J. Phys. C {\bf 7}, 1046 (1974).

\bibitem{JKKN-77} J. V. Jos\'e, L. P. Kadanoff, S. Kirkpatrick, and
  D. R. Nelson, Renormalization, vortices, and symmetry-breaking
  perturbations in the two-dimensional planar model, Phys. Rev. B
  {\bf 16}, 1217 (1977).

\bibitem{PV-13} A.Pelissetto and E. Vicari, Renormalization-group flow
  and asymptotic behaviors at the Berezinskii-Kosterlitz-Thouless
  transitions, Phys. Rev. E {\bf 87}, 032105 (2013).

\bibitem{BPV-20-on3} C. Bonati, A. Pelissetto, and E. Vicari,
  Three-dimensional phase transitions in multiflavor scalar SO($N_c$)
  gauge theories, Phys. Rev. E {\bf 101}, 062105 (2020).

\bibitem{PV-19} A.~Pelissetto and E.~Vicari, Multicomponent compact
  Abelian-Higgs lattice models, Phys. Rev. E {\bf 100}, 042134 (2019).

\bibitem{BPV-19-qcd3} C. Bonati, A. Pelissetto and E. Vicari, Phase
  diagram, symmetry breaking, and critical behavior of
  three-dimensional lattice multiflavor scalar chromodynamics,
  Phys. Rev. Lett. {\bf 123}, 232002 (2019).

\bibitem{BPV-20-qcd3} C. Bonati, A. Pelissetto and E. Vicari, Phase
Three-dimensional lattice
  multiflavor scalar chromodynamics: interplay between global and
  gauge symmetries, Phys. Rev. D {\bf 101}, 034505 (2020).

\bibitem{CEPS-93}
S. Caracciolo, R. G. Edwards, A. Pelissetto, and A. D. Sokal,
New universality classes for two-dimensional $\sigma$-models,
Phys. Rev. Lett. {\bf 71}, 3906 (1993); 
S.~Caracciolo, A.~Pelissetto and A.~D.~Sokal,
Analytic Results for Mixed $O(N)$/$RP^{N-1}$ $\sigma$-Models in
Two Dimensions,
Nucl. Phys. {\bf 34} (Proc. Suppl.), 683 (1994).

\bibitem{Hasenbusch-96}
M. Hasenbusch,
O($N$) and RP$^{N-1}$ models in two dimensions,
Phys. Rev. D {\bf 53}, 3445 (1996).

\bibitem{NWS-96}
F. Niedermayer, P. Weisz, and D.-S. Shin,
Question of universality in RP$^{N-1}$ and O($N$) lattice
$\sigma$ models,
Phys. Rev. {\bf 53}, 5918 (1996).

\bibitem{CHHR-98}
S. M. Catterall, M. Hasenbusch, R. R. Horgan, and R. Renken,
Nature of the continuum limit in the 2D $RP^2$ gauge model,
Phys. Rev. D {\bf 58}, 074510 (1998).

\bibitem{BFPV-20}
C. Bonati, A. Franchi, A. Pelissetto, and E. Vicari,
Asymptotic low-temperature behavior of two-dimensional
RP$^{N-1}$ models,  {\tt arXiv:2006.13061}.

\bibitem{FB-72} M. E. Fisher and M. N. Barber, Scaling theory for
  finite-size effects in the critical region, Phys. Rev. Lett. {\bf
    28}, 1516 (1972).

\bibitem{Barber-83}
  M. N. Barber, in {\em Phase Transitions and Critical Phenomena},
  edited by C. Domb and J. L. Lebowitz (Academic Press, New York, 1983),
  Vol. 8.

\bibitem{Privman-90} V. Privman ed., {\em Finite Size Scaling and
  Numerical Simulation of Statistical Systems} \/ (World Scientific,
  Singapore, 1990).

\bibitem{PV-02} A. Pelissetto and E. Vicari, Critical phenomena and
  renormalization group theory, Phys. Rep. {\bf 368}, 549 (2002).

\bibitem{CP-98} 
S. Caracciolo and A. Pelissetto, Corrections to
finite-size scaling in the lattice $N$-vector model for $N=\infty$,
Phys. Rev. D {\bf 58}, 105007 (1998).

\bibitem{BNW-10}
J. Balog, F. Niedermayer, and P. Wiesz,
The puzzle of apparent linear lattice artifacts in the 2d non-linear
sigma-model and Symanzik's solution,
Nucl. Phys. B {\bf 824}, 563 (2010).

\end{thebibliography}
\end{document}